\documentclass[12pt]{iopart}

\usepackage{graphicx}

\begin{document}

\title[Tuning the effective coupling of an AFM lever to a thermal bath]{Tuning the effective coupling of an AFM lever to a thermal bath}

\author{G Jourdan$^{1,4}$, G Torricelli$^2$, J Chevrier$^{1,3}$ and F Comin$^3$}

\address{$^1$ Institut N\'eel CNRS Grenoble BP 166 38042 Grenoble cedex 9 France and Universit\'e Joseph Fourier - BP 53 - 38041
Grenoble Cedex 9}
\address{$^2$Department of Physics and Astronomy, University of Leicester, University Road, Leicester, LE1 7RH}
\address{$^3$ ESRF, 6 rue Jules Horowitz, BP220, 38043 Grenoble Cedex, France}
\address{$^4$ Laboratoire Kastler Brossel, CNRS Paris, ENS, UPMC, 4 Place Jussieu, 75252 Paris Cedex 05}

\ead{guillaume.jourdan@grenoble.cnrs.fr}

\begin{abstract}
Fabrication of Nano-Electro-Mechanical-Systems (NEMS) of high quality is nowadays extremely efficient. These NEMS will be used as sensors and actuators in integrated systems. Their use however raises questions about their interface (actuation, detection, read out) with external detection and control systems. Their operation implies many fundamental questions related to single particle effects such as Coulomb blockade, light matter interactions such as radiation pressure, thermal effects, Casimir forces and the coupling of nanosystems to external world (thermal fluctuations, back action effect). Here we specifically present how the damping of an oscillating cantilever can be tuned in two radically different ways: i) through an electro-mechanical coupling in the presence of a strong Johnson noise, ii) through an external feedback control of thermal fluctuations which is the cold damping closely related to Maxwell's demon. This shows how the interplay between MEMS or NEMS external control and their coupling to a thermal bath can lead to a wealth of effects that are nowadays extensively studied in different areas.
\end{abstract}

\vspace{2pc}
\submitto{Nanotechnology}
\maketitle

\section{Introduction}

\subsection{General context}
NEMS with dimensions truly nanometric are fascinating objects. They are now described in numerous reviews \cite{ekinci:061101,Roukes:2001:PRI} and are fascinating in many respects. They are geometrically and chemically very well defined model objects, with numerous potential applications. Well known examples are oscillators such as silicon levers or carbon nanotubes. As NEMS contain a large number of atoms (usually much more than a million), they are most often described as continuous objects but their linear dimensions can make them geometrically close to complex molecules and to viruses. They are used for example as accelerometers, mass detectors with sensitivity close to the atomic level \cite{ekinci:4469}, and electro-mechanical filters at high frequencies. The description and analysis of their properties inherently touch some very fundamental issues related to electron or photon coupling with a mechanical oscillator. Driving effects in this field are certainly the Casimir force due to the zero point motion of the electro-magnetic field in vacuum, and observation of the quantum behavior of a NEMS oscillator \cite{schwab:36,M.D.LaHaye04022004,Arcizet,Klechner,Gigan}. It is interesting to notice that most groups involved in this research are readers of Braginsky's pioneering works related to measurements of ultra small displacements often in the context of gravitational wave detection \cite{Braginsky1,Braginsky2}. Small energies involved in NEMS excitations make them very sensitive to many external noises. Specifically one can mention measurement back action effects and Brownian motion as they are always operated in contact with a thermal bath. Due to thermal fluctuations, it is hardly possible to define true quasi-static transformations as these NEMS are driven from one stable state to another. In the last decade, in the context of molecular motors, this point has been addressed both theoretically and experimentally. It shows that far before touching any quantum limit, thermal fluctuations that are usually negligible in mechanics dealing with current objects (although not in electronic with strong Johnson noise often limiting performances), become a central ingredient in the description of NEMS behaviour. A decade ago, general theorems in statistical physics \cite{PhysRevLett.78.2690,Crooks} such as the Jarzynski equality, an equation that connects the free energy difference between two equilibrium states and the work needed to travel between these two states through nonequilibrium processes, and recently experimentally studied \cite{douarche-2005}, have given a frame to analyse the non quasi-static processes induced by thermal fluctuations that are involved in NEMS operation at room temperature. This has already been applied to AFM experiments \cite{kim-2007-75,noy:4792}. In this paper, we show how it is possible to alter the resonance curve width of an oscillator coupled to a thermal bath at temperature T. We then present the result of two independent experiments performed using a UHV Omicron AFM and a home made AFM. The two examples presented here have been chosen so that in one case it is the coupling to the thermal bath that has been directly changed tuning the electro-mechanical coupling, whereas in the second case, this coupling has remained the same but a very fast control of the thermal fluctuations has been set resulting in a very efficient thermal energy extraction from the oscillator.

\subsection{Report on two AFM experiments}
 Experimental report that thermally induced fluctuations of a mechanical nano or micro-oscillator can be directly and efficiently influenced without changing the bath temperature is then the central point of this paper. To exemplify this point, we present two different AFM experiments. AFM levers are used as a model system for MEMS and NEMS. Although the oscillator has micrometer dimensions, displacements and distances of interaction are truly nanometric. Both experiments have in common the experimental control of the thermally induced fluctuations of the mechanical oscillator connected to a
thermal bath. The signature of the dissipative coupling efficiency onto the resonance of a mechanical oscillator is as usual identified in the frequency resonance width. In the first case, a true increase of the dissipation is observed due to an added dissipative channel as the oscillator becomes part of an electrical circuit that includes a large resistance. In the second case in contrast, an apparent increase of dissipation is in fact due to detailed damping of thermal fluctuations by an external feedback loop. The resonance width clearly increases but a simultaneous and large decrease of oscillator thermal energy is observed and is analysed as a strong reduction in temperature.

\section{AFM: increase damping due to Johnson noise} 

\begin{figure}[h]
	\centering
		\includegraphics[width=1.00\textwidth]{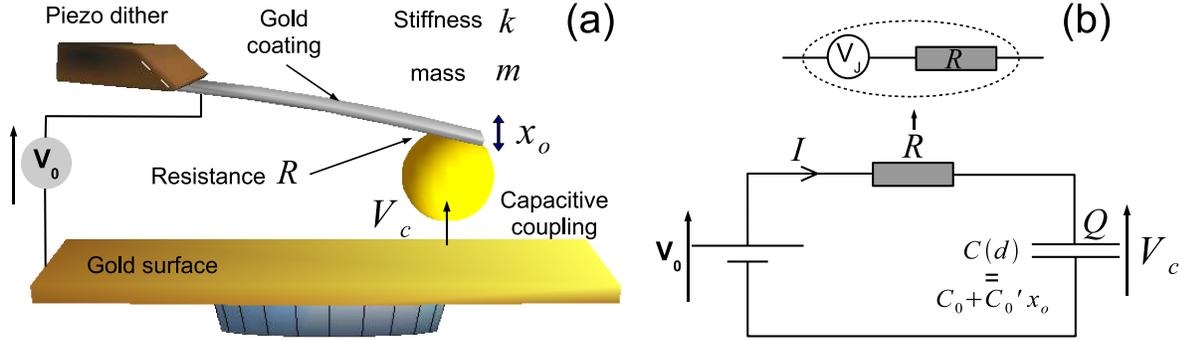}
	\caption{(a) The metallic sphere radius $R_s$ is about $40 \; \mu \mathrm{m}$. The sphere surface distance $d$ is between $50$ nm and $500$ nm. A voltage is established between the sample surface and the sphere: both form a distance dependent capacitance $C(d)$.  The resistance R is close to $ 30 \; \mathrm{M} \Omega$ and is due to a poor metallic coating of the AFM lever. (b) Schematic of the polarization circuit: it consists of a resistance and capacitance device. The capacitance depends on the lever motion through $C(d)=C_0+C'_0 x_o$. The inset above illustrates the Johnson model of a resistance: a perfect resistance associated with a noisy voltage source $V_{\mathrm{J}}$.}
	\label{fig:figure1_a_b}
\end{figure}

As classically modelled by the Langevin equation, coupling of AFM levers or of micro/nanooscillators to a thermal bath is essentially described by a viscous force $-\Gamma v$ in the dynamical equation. A characteristic time $\tau = m/ \Gamma$ can be associated with the friction coefficient $\Gamma$, where $m$ is the oscillator mass. The oscillator resonance frequency $\omega_{res}$ is much larger than $1/ \tau$. Short time scale thermal fluctuations with a known force spectral density are described by the temperature and the same friction coefficient $ \Gamma$ through a constant spectral density of noise:
\begin{equation}
S_{F} = 2 k_{B} T \Gamma
\label{eq:f_noise_density}
\end{equation}
The time $\tau=m/ \Gamma$ as discussed in reference [9] characterizes the long time oscillator behaviour and the kinetic of heat exchanges with the thermal bath. For time $t<<\tau$,  the oscillator keeps the memory of its state at $t=0$, whereas for time $t>>\tau$,  this memory is lost and the average oscillator displacement is fixed by the thermal bath and described by the fluctuation-dissipation theorem (energy equipartition).
Measurement of $\tau$ from the oscillator resonance characteristics, essentially from the resonance width, enables us to investigate specific channels of energy dissipation in the thermal bath. We here present the study of a dissipation channel associated with Johnson noise present in the set up described in figure~\ref{fig:figure1_a_b}.

The set up consists of a gold coated sphere with a radius $R_{\mathrm{s}} = 40 \; \mu \mathrm{m}$ glued at the free end of an AFM cantilever creating a capacitance with the sample as the opposite electrode (figure~\ref{fig:figure1_a_b}). The microsphere polarization circuit is controlled by a DC generator and includes a large  resistance due to a poor metallic coating of the probe (this resistance is shown in following section to be $30 \; \mathrm{M} \Omega$). When applying a DC voltage $V_{\mathrm{c}}$, an attractive capacitive force between the sphere and the sample surface is then generated and described by: 
\begin{equation}
F_e= \frac{1}{2} C'_0 V_{\mathrm{c}}^{2}
\label{eq:capacitive_force}
\end{equation}
In the sphere plane configuration, the derivative of the capacitance is given at small distances ($d<<R_{\mathrm{s}}$) by:
\begin{equation}
C'_0 = - \frac{2 \pi \epsilon_0 R_{\mathrm{s}}}{d}
\label{eq:sphere_plane_d_capacitance}
\end{equation}
This electromechanical coupling turns out to be a new damping channel for the main degree of freedom $x_{o}$ of the mechanical oscillator, since the latter is now connected to the thermal bath associated to the resistance.

\subsection{The damping process}
In this section we show that the mechanical energy can be dissipated by the joule effect in addition to other mechanical friction processes described by $\Gamma_0$ (phonons collisions, etc.).
The potential between the sample and the sphere amounts to $V_{\mathrm{c}}=V_0-RI$ (figure~\ref{fig:figure1_a_b}(b)). Therefore, to first order in $I$, a contribution to the force directly proportional to the current is generated. According to the equation~(\ref{eq:capacitive_force}) this contribution is given by: 
\begin{equation}
F_e= - C'_0 V_0 RI
\label{eq:current_force}
\end{equation}
For small displacement in first order of $x_{0}$ and $I$:
\begin{equation}
Q=C(d_0+x_{o}) V_{\mathrm{c}} = (C_0 + C'_0 x_{o})V_0 - R C_0 I
\label{eq:Q_charge}
\end{equation}
The charge $Q$ located on the sphere depends on the lever motion $x_{o}$ because of the electromechanical coupling. The last term can be neglected when compared to Q, because the working frequency range determined by the resonance ($\omega_{res}/ 2 \pi = 5009.8 \;\mathrm{Hz}$) is well below the cut frequency $1/RC_0$ (at 100 nm, a pessimistic estimation gives 12000 Hz if the sphere is modelled by a flat 50 $\mu$m radius disk) of the low pass circuit. It results a current proportionnal to the speed of the lever according to the relation $I=dQ/dt=C'_0 \dot{x_{o}} V_0$, the force thus obtained turns out to be a damping process:  
\begin{equation} 
F_e = - (C'_{0})^{2} V_{0}^{2} R \dot{x_{o}}
\label{eq:Johnson_dissipation_force}
\end{equation}
Therefore the resistance transforms the current generated by the lever displacement into heat with a friction coefficient $\Gamma_R = (C'_{0})^{2} V_{0}^{2} R$.
\begin{figure}[h]
	\centering
	\includegraphics[width=0.49\textwidth]{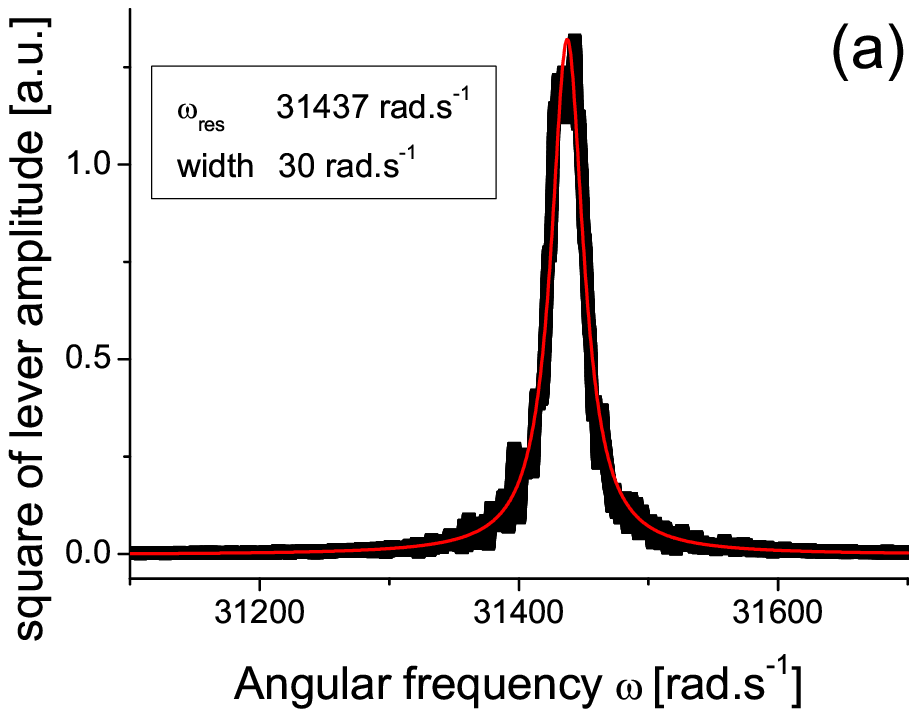}
	\includegraphics[width=0.49\textwidth]{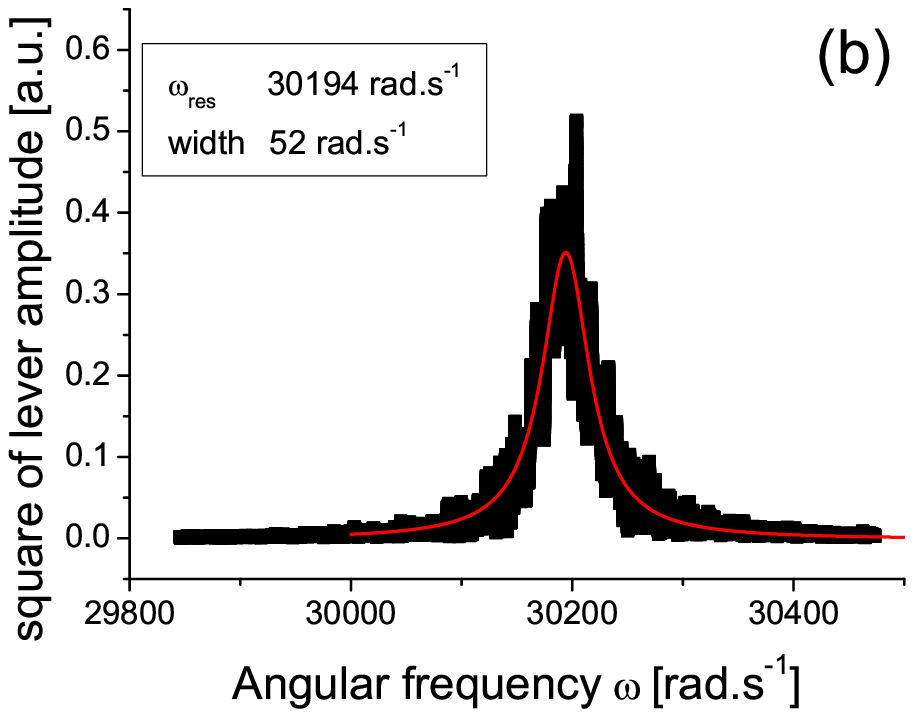}
	\caption{Lever resonance curves (fundamental mode of vibration) are shown for two different coupling voltages.(a) no voltage is applied $V=0.0$ V. (b) a voltage is applied $V=0.5$ V. Sphere surface distance is here about $90$ nm. Conclusion: Applying $0.5$ V induces a width increase close to a factor 2. This is a fully reversible and controllable process. The electrostatic force gradient yields the resonance frequency decrease of the second curve.}
	\label{fig:figure2a}
\end{figure}
Experimentally an increase of the dissipation rate is clearly observed when a bias voltage is applied to the system. As shown in figure~\ref{fig:figure2a}, at a distance of 90 nm width increase from 30 rad/s up to 52 rad/s is observed in the resonance curve when the applied voltage is changed from 0 to 0.5 V. This is a clear evidence that the increased dissipation has an electrostatic origin. It implies therefore the existence of a dissipative channel that is made efficient as the sphere surface voltage is applied in agreement with equation~ (\ref{eq:Johnson_dissipation_force}). It means that this added dissipative channel is externally tuned by the applied voltage $V_0$. 

The effective friction coefficient $\Gamma_R = m \gamma_R$ can be used to analyse the continuous increase of the resonance width as the sphere surface distance $d$ is decreased at constant applied voltage of $0.5 \; V$. The behaviour of the width parameter $\gamma = \gamma_0 + \gamma_R$ is consistent with a $1/d^2$ power laws predicted by equations (\ref{eq:sphere_plane_d_capacitance}) and ({\ref{eq:Johnson_dissipation_force}):
\begin{equation}
\gamma_R = \frac{4 \pi^2 \epsilon_{0}^{2} R_{s}^{2}}{d^2}\frac{V_{0}^{2} R}{m}
\label{eq:gamma_R}
\end{equation}
The mass of the oscillator is dominated by the microsphere and estimated to $m = 2.8 \; 10^{-10}$ kg ($\rho_{sphere}=1040 \; \mathrm{kg}\mathrm{m}^{-3}$).
\begin{figure}[h]
	\centering	\includegraphics[width=0.50\textwidth]{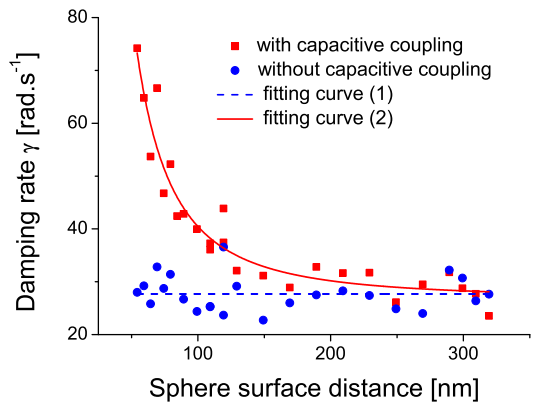}
	\caption{Widths deduced from resonance curves like ones in figure~\ref{fig:figure2a} are reported for different sphere plane distances. Curves are shown for two voltages: blue dots, V=0.0 volt; red squares V=0.5 volt. The fitting curve (1) determines the average value of the mechanical friction coefficient $\gamma_0 = 27.7 \pm 0.7 \; \mathrm{s}^{-1}$. The fitting curve (2) relies on equation~(\ref{eq:gamma_R}): it is found that $4 \pi^2 \epsilon_{0}^{2} R_{s}^{2} V_{0}^{2}/ m = 1,4 \; 10^{-13}$ SI. It means that a $30 \; \mathrm{M} \Omega$ resistance can take into account this phenomenon. }
	\label{fig:figure3}
\end{figure}
In figure~\ref{fig:figure3}, a resistance of $30 \; \mathrm{M} \Omega$ is used to quantitatively describe the experimental result. This is the single adjustable parameter as all other quantities are experimentally measured in this model geometry.

\subsection{The fluctuation dissipation theorem: thermal equilibrium}
The new dissipative channel can then be externally controlled as the electromechanical coupling is made efficient using a circuit that includes a dissipative element such as a resistance. However this increased damping does not mean that mechanical position fluctuations are more damped than without any coupling. Conversely the Johnson noise $V_{\mathrm{J}}$ introduced by the resistance in the circuit above turns out to be a new source of thermal excitation of the lever (figure~\ref{fig:figure1_a_b}). It indeed produces an additional fluctuating force on the sphere through the crossing term $2 V_0 V_{\mathrm{J}}$:
\begin{equation}
F_{\mathrm{J}} = C'_0 V_0 V_{\mathrm{J}}
\label{eq:johnson_force}
\end{equation}
whose power density spectrum is given by  $S_{F_{\mathrm{J}}}= (C'_0)^2 V_{0}^{2} S_{V_{\mathrm{J}}}$ with $S_{V_{\mathrm{J}}} = 2 k_{\mathrm{B}} T R$. Although not original, it remains interesting for consistency of the analysis to show that the results comply with the fluctuation dissipation theorem which establishes that the damping rate $\Gamma$ is related to the Langevin fluctuation force~$F_{\mathrm{L}}$:
\begin{equation}
\Gamma = \frac{1}{2 k_{\mathrm{B}} T} \int < F_{\mathrm{L}}(t) F_{\mathrm{L}}(0)> dt
\label{eq:force_Langevin}
\end{equation}
For the noisy force $F_{\mathrm{J}}$, this comes from equation~(\ref{eq:johnson_force}) and $S_{V_{\mathrm{J}}} = \int V_{\mathrm{J}}(t) V_{\mathrm{J}}(0) dt$:
\begin{equation}
\Gamma_{\mathrm{J}} = (C'_0 V_0)^2 R
\label{eq:johnson_damping_rate}
\end{equation}
This is the rate $\Gamma_R$ derived in equation~(\ref{eq:Johnson_dissipation_force}). Since the oscillator is in thermal equilibrium with its surroundings, the fluctuation intensity as well as friction rate gets higher. Energy exchanges between the microlever and the thermal bath increase, but the balance remains the same: the overall flux is nil. The equipartition theorem provides therefore the average energy of the system $\frac{1}{2} k <x_{o}^{2}> = \frac{1}{2} k_{\mathrm{B}} T$.
This experimental example shows how it is possible to control a dissipative coupling between a micromechanical oscillator and a thermal bath. Here an electro-mechanical coupling is established and dissipation takes place as currents induced by oscillator vibrations are dissipated in the resistance R also at the origin of Johnson noise shown in equation~(3). The coupling between the thermal bath and the oscillator is then truly tuned. Finally, it is important to note that the whole system remains in thermodynamic equilibrium with the thermal bath: the energy exchanges is balanced. This is in contrast with experimental strategy explored in next section.

\section{AFM and cold damping}

As indicated by the title of the pioneering article by Milatz, Van Zolingen and Van Iperen \cite{MZI} in $1953$, "The reduction in the Brownian motion of electrometers", it is possible to damp thermal fluctuations. However this cannot be done through a simple increase of the friction coefficient as this would immediately increase the thermal noise spectral density of force (equation~(\ref{eq:f_noise_density})) with a cancellation of both effects to maintain fluctuations on position fixed by the equipartition principle of energy at thermal equilibrium.

It has been demonstrated in several circumstances that using an external feedback loop that acts against thermal fluctuations with an applied force proportional to the oscillator speed, it is really possible to damp these fluctuations. This is the so called cold damping procedure \cite{PhysRevLett.83.3174,PhysRevA.63.013808,PhysRevB.68.235328}.

In the case of cold damping experiment based on an AFM set up, an active cooling can be implemented using an external laser applied onto the lever \cite{mertz:2344}, and either using a capacitive coupling or a bimorph piezo actuator, that is normally used to excite the lever at resonance, to rapidly react against the thermal fluctuations. We have successfully used the two latter strategies. We shall here describe the third one usable on any AFM. Compared to capacitive coupling it seems to be less intuitive. The experimental set-up is schematically described in figure~\ref{fig:figure4}.

\begin{figure}[h]
	\begin{center}
		\includegraphics[width=0.50\textwidth]{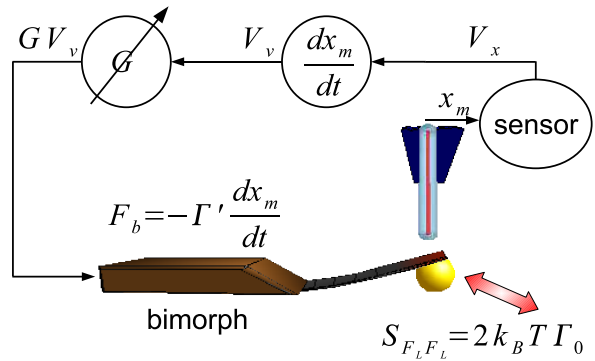}
	\end{center}
	  \caption{Schematic description of the cold damping feedback loop. The microlever is damped by the thermal bath with a rate $\Gamma_0$ and by the feedback loop through the bimorph actuator with a rate $\Gamma_r$: The overall friction coefficient is $\Gamma = \Gamma_0 + \Gamma_r$. As the loop gain $G$ is increased, the
damping becomes more and more efficient as it is shown in the figure~\ref{fig:figure5}.}
	\label{fig:figure4}
\end{figure}

The measured lever position $x_{o}$ is directly used through a feedback loop to apply a damping force $F_D$ proportional to the lever velocity, i.e. the $x_{o}$ time derivative, that counteracts the displacement. The ratio between the damping force $F_D$ and the lever speed is $\Gamma_r$. The measured spectral density of the lever displacement is shown in figure~\ref{fig:figure5} as $\Gamma_r$ is increased. As it is well known when cold damping is applied to an oscillator, the peak width is increased, and therefore the apparent friction coefficient is also increased. However the peak area decreases gradually which is classically described as a temperature reduction for the vibration mode. These new characteristics can be compared to the friction coefficient $\Gamma_0$ and the temperature $T$ when no active loop is present. The characteristic time $\tau_0 = m/ \Gamma_0$ defined with no active loop becomes $\tau_r = m/ (\Gamma_0 + \Gamma_r)$ with cold damping operating. As the spectral density of force remains the same with and without cold damping operating, neither the thermal bath nor the system coupling to thermal bath are affected, and the spectral density of force remains the same whatever the value of $\Gamma_r$:
\begin{equation}
S = 4 k_B T \Gamma_0 = 4 k_B T_{eff} (\Gamma_0+\Gamma_r)  
\end{equation}
This equality \cite{PhysRevLett.78.2690} leads to the following expression for the resonance mode temperature:
\begin{equation}
T_{\mathrm{eff}} = T \frac{\Gamma_0}{\Gamma_0+\Gamma_r}
\label{eq:T_eff}
\end{equation}
\begin{figure}[h]
	\centering
	\includegraphics[width=0.49\textwidth]{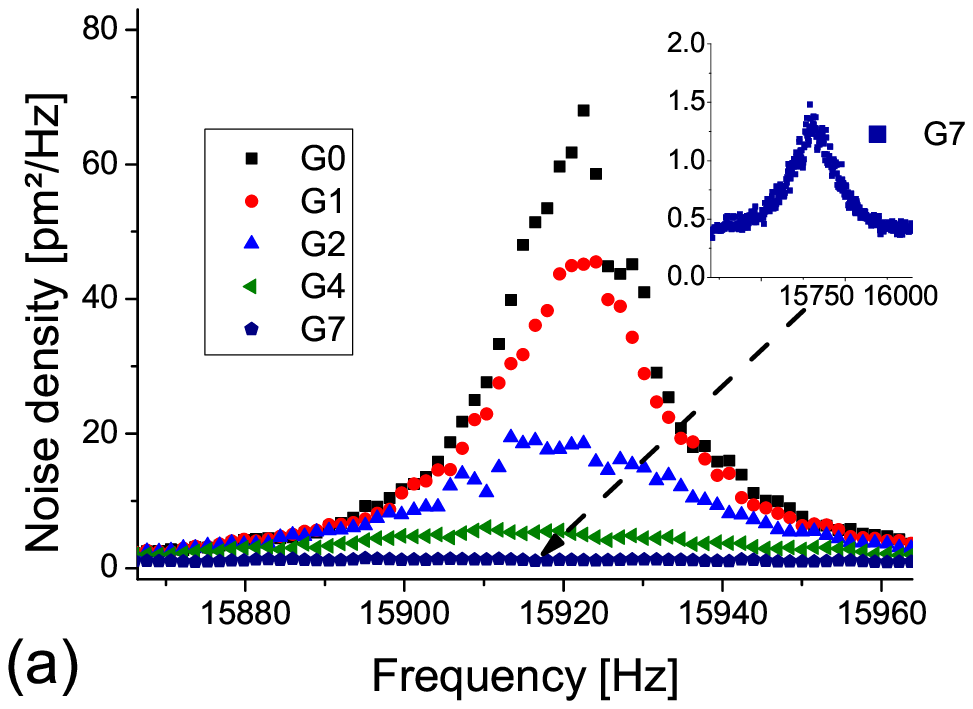}
	\includegraphics[width=0.49\textwidth]{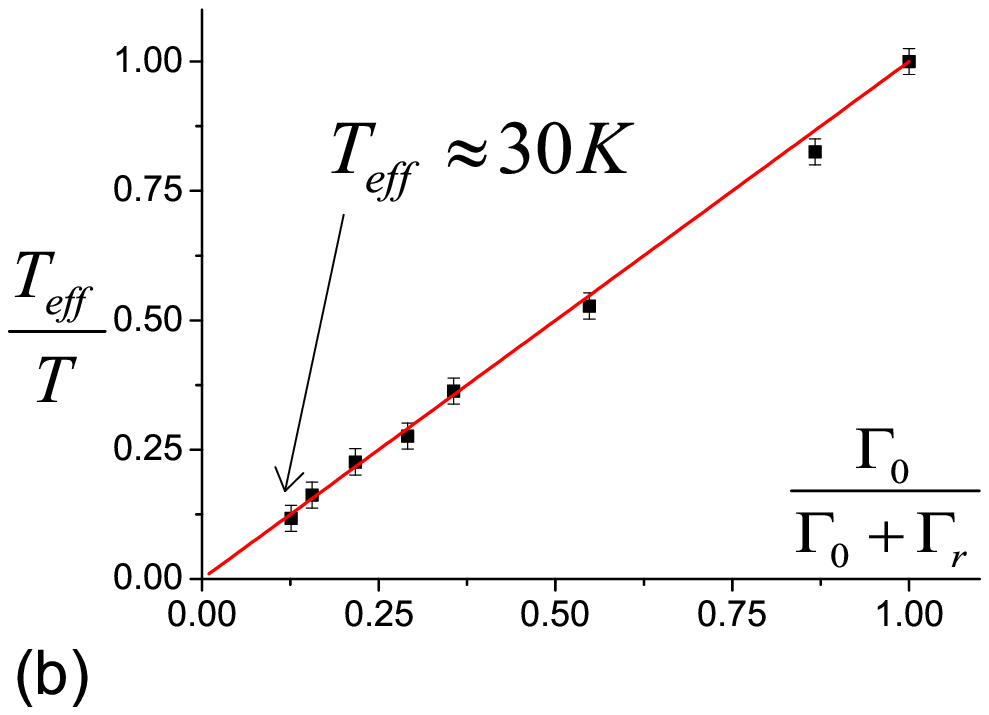}
	\caption{(a) Displacement spectral density: the thermal fluctuations are progressively cancelled as the loop gain $G$ is increased from G0 to G7 (figure~\ref{fig:figure4}). The large thermal displacement at resonance has almost vanished. The small resonance frequency shift is due to a remaining proportional gain in the feedback loop device. (b) The effective temperature is proportional to the area under each lorentzian curve, namely the motion dispersion $<x_{o}^{2}>$. As the feed back loop system is implemented, it decreases from room temperature down to about $30$ K.}
	\label{fig:figure5}
\end{figure}

The thermal fluctuations measured at frequencies around this resonance are then progressively decreased as the feedback loop gain is increased. A quantitative and detailed analysis of this cold damping procedure shows that as the loop gain is increased, the peak integral is strongly reduced and the apparent resonant mode temperature is strongly decreased from 300 K down to 30 K. This is an apparent temperature as this is by no means a thermal bath temperature but the result of the stationary oscillator energy exchange with the thermal and with the external applied force. This we shall analyze below.

\section{Conclusion: comparison of these two experiments}

In both experiments presented here, the resonance curve is broadened which is immediately analyzed as an increase of the effective friction coefficient. This is a common point between the two experiments. However a discussion of the thermal energy transfer in both cases shows how much these two procedures differ.

In the first case, adding a capacitive coupling in the presence of a large resistance does not introduce a qualitative change. The friction coefficient is increased and the oscillator coupling to thermal bath is made more efficient with a shorter time $\tau$. This added coupling induces a broadening of the resonant peak but the integral of the resonance peak is essentially preserved and fixed by the thermal bath temperature. The thermal noise spectral density of force increases accordingly. The major change is then an increased coupling of the oscillator to the thermal bath with consequently faster heat
exchanges but also a larger thermal noise spectral density of force applied to the oscillator. Originality here relies in the fact that the increased dissipation is due to the charge fluctuation at the sphere/surface due to the Johnson noise of the resistance.
In the second case, the resonant peak also broadens. Although this is an apparent increase of the friction coefficient, this is not due to an increased direct coupling between the oscillator and the thermal bath. This coupling to the thermal bath does not change as the feedback loop is activated to induce cold damping. A new coupling is introduced with an external "intelligent" system which very rapidly and efficiently extracts thermal energy from the oscillator. This real time precise and detailed action directly on the thermal fluctuations is very much reminiscent of the way a Maxwell demon is acting as noticed in \cite{kim-2007-75} that provides a theoretical description of this type of experiment related to general theorems \cite{PhysRevLett.78.2690,Crooks}. In practice this extraction is so fast and so efficient that there is no time for the heat bath to impose its own temperature of 300 K. Thermal fluctuations are damped in detail and the temperature subsequently decreases. As this is a fast process of thermal energy extraction, the measured characteristic time $\tau_r$ on the resonance curve is now much shorter than $\tau_0$ the characteristic time of heat exchange when there is no cold damping.

It is then interesting to notice that it is the high quality factor of the oscillator i.e. the low friction coefficient that prevents the heat bath to efficiently maintain the energy oscillator at the value of $k_B T$ when the cold damping is acting. In stationary regime, the temperature determined from the oscillator fluctuation distribution is much smaller than the heat bath temperature, i.e. the cold damping process finally extracts energy from the heat bath through the oscillator. The whole process can then be described as a « microscopic fridge» as quoted in \cite{kim-2007-75}. However an estimate of involved heat fluxes shows that it is not easy to make this fridge efficient enough to really cool down a thermal bath, even if it is a non realistic isolated nanosystem.

As a final remark, regarding NEMS sensor resolution, two different outcomes can be drawn from these two damping processes. In both cases, the mechanical response function of the lever is tuned as a result of the introduction of a new force on the system. However, force measurement accuracy is at best set by the thermal bath fluctuation level. The Johnson noise coupling is then undesirable for applications such as an accelerometer, a tiny force sensor etc. Therefore one has to pay attention to this additional damping process when implementing a capacitive coupling in a nanosystem. In contrast, the cold damping technique does not affect the force measurement noise. However, this comment remains rigorously true if the outside control system can be considered as perfect. Detection noise indeed limits cold damping efficiency and will be studied in a further article.

\begin{figure}[h]
	\centering	     	  
	\includegraphics[height=4cm]{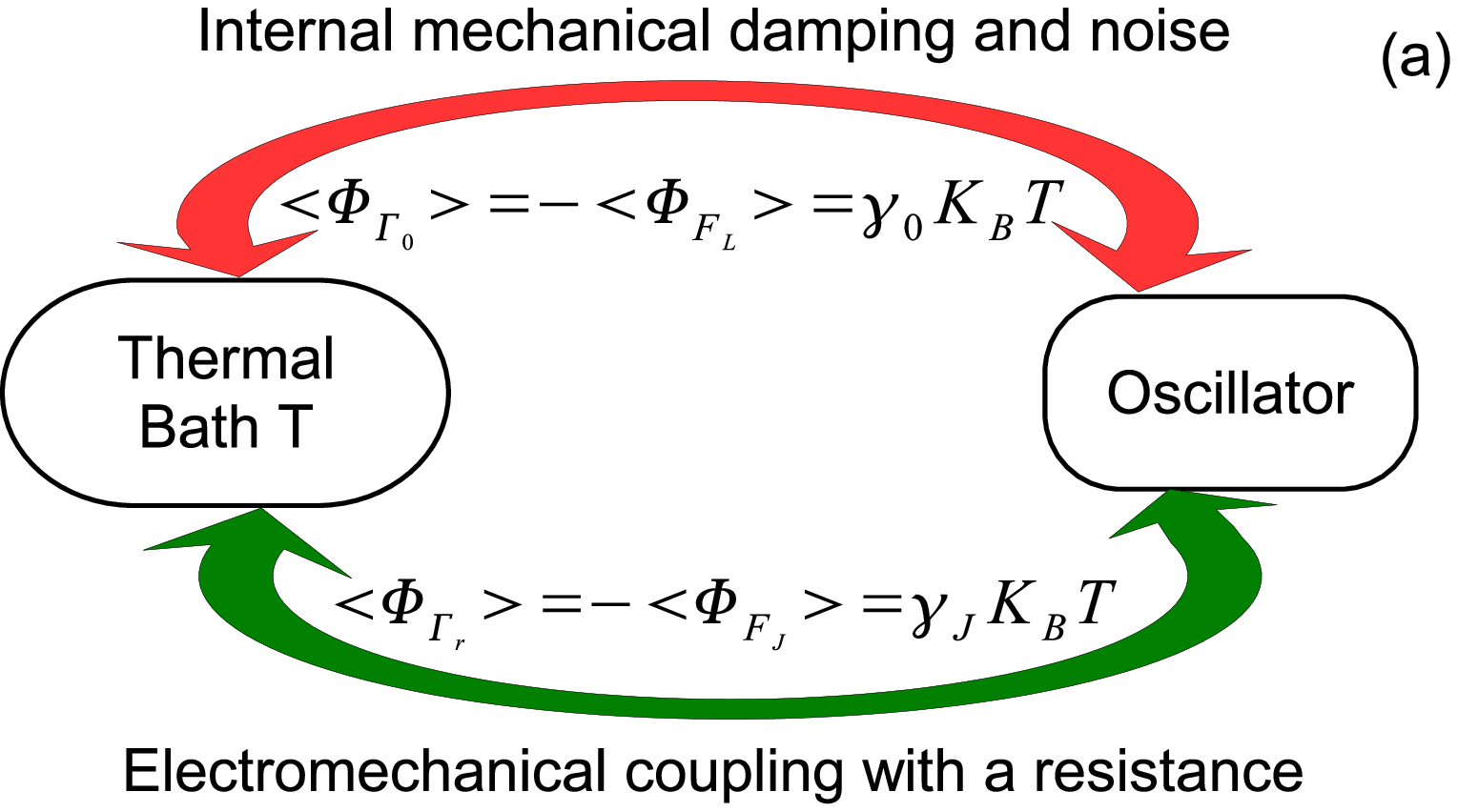}
	\includegraphics[height=4cm]{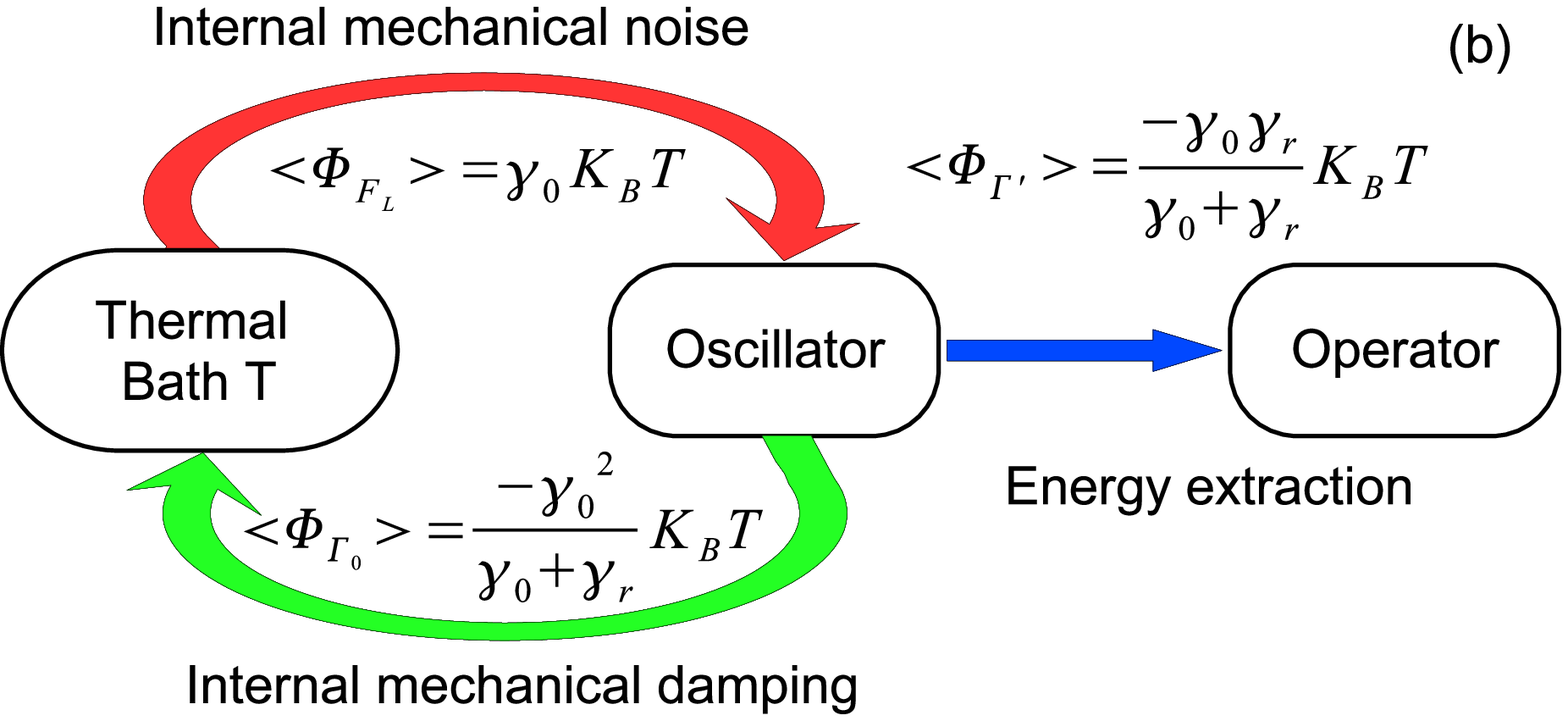}
		\caption{(a) The capacitive coupling adds a classical dissipative channel due to an electromechanical coupling that couples the Johnson noise to the oscillator. The damping is increased but thermal fluctuations are accordingly increased. (b) The cold damping feedback loop extracts thermal energy from the oscillator by direct damping of thermal fluctuations. The damping is increased but no energy is randomly provided by the feedback loop to the oscillator. The temperature is decreased as thermal energy is extracted much more rapidly than what can be provided by the thermal bath.}
	\label{fig:figure6a}
\end{figure}

\section*{References}
\bibliographystyle{unsrt} 
\bibliography{biblio} 

\end{document}